\documentclass[aip,jcp,numerical,preprint]{revtex4-2}

\usepackage{amsmath}
\usepackage{epstopdf}
\usepackage{graphicx}
\usepackage{dcolumn}
\usepackage{bm}
\usepackage{braket}
\usepackage{gensymb}
\usepackage{array}
\newcolumntype{P}[1]{>{\centering\arraybackslash}p{#1}}
\newcolumntype{M}[1]{>{\centering\arraybackslash}m{#1}}

\begin{document}

\title{Theory of the Dark State of  Polyenes and Carotenoids}

\author{William Barford}
\email{william.barford@chem.ox.ac.uk}
\affiliation{Department of Chemistry, Physical and Theoretical Chemistry Laboratory, University of Oxford, Oxford, OX1 3QZ, United Kingdom}

\begin{abstract}
A theory is developed to describe the singlet dark state (usually labeled $S_1$ or  $2A_g$) of polyenes and carotenoids. The theory assumes that in principle this state is a linear combination of a singlet triplet-pair and an odd-parity charge-transfer exciton. Crucially, these components only couple when the triplet-pair occupies neighboring dimers, such that an electron transfer 
between the triplets creates a nearest-neighbor charge-transfer excitation. This  local coupling stabilises the $2A_g$ state and induces a nearest neighbor attraction between the triplets. In addition, because of the electron-hole attraction in the exciton, the increased probability that the electron-hole pair occupies neighboring dimers enhances the triplet-triplet attraction: the triplet pair is `slaved' to  the charge-transfer exciton. The theory also predicts that as the Coulomb interaction is increased, the $2A_g$ state evolves from a predominately odd-parity charge-transfer exciton state with a small component of triplet-pair character to a state predominately composed of a triplet-pair with some exciton character. Above a critical Coulomb interaction  there is a decoupling of the triplet-pair and charge-transfer exciton subspaces, such that the $2A_g$ state becomes entirely composed of an unbound spin-correlated triplet pair.
The predictions of this theory are qualitatively consistent with high-level density matrix renormalization group calculations of the Pariser-Parr-Pople (or extended Hubbard) model.
\end{abstract}

\maketitle


\section{Introduction}\label{Se:1}

The low-energy singlet dark state of polyenes, usually labeled $S_1$ or  $2A_g$, has continued to fascinate researchers for over 50 years\cite{Hudson72,Schulten72}. Its intriguing electronic properties are a consequence of electron-electron interactions and electron-nuclear coupling\cite{Book}.

The triplet-pair or bimagnon character of the  $2^1A_g$ state was first predicted theoretically by Schulten and Karplus in 1972\cite{Schulten72} and further elucidated by Tavan and Schulten in 1987\cite{Tavan87}. Using the Hubbard-Peierls model, Hayden and Mele\cite{Hayden86} then demonstrated the combined role of electronic interactions and electron-nuclear coupling in determining the four-soliton structure of this state for 16 C-atom chains. The four-soliton structure was further investigated for polyene chains of up to 100 C-atoms by solving the Pariser-Parr-Pople-Peierls (PPPP) model using the density matrix renormalization group (DMRG) method\cite{Bursill1999,Barford01}. Since a triplet excitation coupled to  the nuclei creates a soliton-antisoliton ($\textrm{S}\bar{\textrm{S}}$) pair, the four-soliton structure (i.e., a pair of soliton-antisoliton pairs) is further strong theoretical evidence of the triplet-pair character of the  $2^1A_g$ state.

Solitons (${\textrm{S}}$) and antisolitons ($\bar{\textrm{S}}$) are associated with domain walls in the bond alternation of linear polyenes\cite{Ooshika1957b,Longuet-Higgins1959}. They are also associated with spinons, the elementary excitations of a one-dimensional spin-1/2 quantum antiferromagnet\cite{Pople1962,Uhrig1996}. A spinon is a charge-neutral, spin-1/2 object, while a triplet excitation is a bound spinon-pair, which is recognized as a $\textrm{S}\bar{\textrm{S}}$ pair in the bond alternation. Thus, the $\textrm{S}\bar{\textrm{S}}$ pair separation is a measure of the internal size of the triplet excitation. Similarly, the $\textrm{S}\bar{\textrm{S}}$-$\textrm{S}\bar{\textrm{S}}$ separation is a measure of the triplet-triplet pair separation. Barford \textit{et al.}\cite{Bursill1999,Valentine20} showed that in polyenes the $\textrm{S}\bar{\textrm{S}}$ separation converges to 4 dimers, while the $\textrm{S}\bar{\textrm{S}}$-$\textrm{S}\bar{\textrm{S}}$ separation converges to 6 dimers, indicating that the triplet-pair is bound. Further numerical evidence that the triplet-pair is bound in the $2^1A_g$ state was provided by Valentine \textit{et al.}\cite{Valentine20}, who showed that its energy converges to a value of $0.3$ eV lower than the energy of a pair of free-triplets.

In the other limit of weak electronic correlations, as in light-emitting polymers, the   $2^1A_g^-$  state is predominately a charge-transfer exciton. The electron-hole wavefunction of this state has odd parity (i.e., the wavefunction is odd under an exchange of the electron and hole), and it lies higher in energy that the optically-allowed $1^1B_u^+$  state (which is usually labeled $S_2$ in polyenes) whose electron-hole wavefunction  has even parity.

The reversal of energies of the $1^1B_u^+$ and  $2^1A_g^-$  states in polyenes is partly a consequence of electronic interactions. As explained later, electronic interactions simultaneously reduce the excitation energy of the lowest covalent triplet state and increase the  excitation energy of ionic singlet states, which implies that for large enough interactions a triplet-pair state will have a lower  energy than a singlet ionic state. The triplet-pair state is further stabilized relative to the ionic state, as electron-nuclear relaxation is enhanced in covalent states.

The energetic reversal of the bright ($S_2$) and dark ($S_1$) states has various photophysical consequences. For example, it explains the non-emissive properties of linear polyenes, it is responsible for the photoprotection properties of carotentoids in light harvesting complexes, and -- because of its triplet-pair character -- it is thought to be the cause of singlet fission in polyene-type systems\cite{Kraabel98,Lanzani99,Musser13,Kasai15,Manawadu2022}.
The electronic states of carotenoids are reviewed in ref\cite{Polivka04,Hashimoto18,Musser19}, while refs\cite{Chan15,Taffet19,Khokhlov20} report on recent high-level \textit{ab initio} calculations of these states.

As already noted, the $2^1A_g$ state of polyenes has been extensively studied via DMRG calculations of the PPPP model. As well as the four-soliton structure, triplet overlaps and spin-spin correlation functions also reveal its triplet-pair character\cite{Valentine20}. Using a suitable exciton creation operator\cite{Barford2008}, Valentine \textit{et al.}\cite{Valentine20} also investigated the excitonic component of this state. Figure 8 of ref\cite{Valentine20} illustrates the odd-parity charge-transfer exciton wavefunction characteristic of  the $2^1A_g$ state of more weakly correlated polymers.

Thus, we can conclude that the $2^1A_g$ state is a linear combination of a singlet triplet-pair and an odd-parity charge transfer exciton. This suggests that the mixing of the triplet-pair and charge-transfer exciton subspaces both stabilizes the $2^1A_g$  state and causes the strong triplet-pair attraction. 
This stabilization is an additional cause of the $1^1B_u$/$2^1A_g$ energy reversal in polyenes.
The theory presented in this paper will explain this property.

The $2^1A_g^-$ state is the lowest energy member of a family of states with the same elementary excitations, but with different pseudomomentum quantum numbers (i.e., $2^1A_g^-, 1^1B_u^-, 3^1A_g^-, \cdots$), which for convenience we label as the `$2A_g$ family'. Irrespective of their even or odd symmetry under a two-fold rotation, these states are all optically dark.
The  singlet triplet-pair component of the $2A_g$ family is optically dark for two reasons. First, it is composed of a pair of electronic excitations and since the dipole operator is a one-electron operator, the transition dipole moment with the ground state vanishes. Second, each electronic excitation is a triplet and since the dipole operator commutes with total spin, the transition dipole moment with the singlet ground state again vanishes. The exciton component of the $2A_g$ family is also optically dark, because its electron-hole wavefunction has odd parity and thus its transition dipole moment with the ground state vanishes\cite{Book}.

In this paper we describe an effective low-energy model of the $2A_g$ family of states in conjugated polyenes that provides a simple, physical explanation of triplet-pair binding. 
There are two key approximations in the model. First, it assumes that polyene chains are comprised of weakly interacting ethylene dimers. This implies that the ground state is composed of a product of singlet dimer states.
Second, it assumes a reduced basis for the excited states, composed of  singlet triplet-pair excitations and odd-parity electron-hole excitations.
The dimer limit implies that the $\textrm{S}\bar{\textrm{S}}$ pair comprising a triplet excitation are bound on a single dimer. As a consequence of these approximations, the theory is only able to provide qualitative predictions for polyene systems, which are nonetheless consistent with DMRG calculations.

Crucially, the theory predicts that when a pair of triplets occupy neighboring dimers an electron transfer  between the dimers connects the pair of triplets to the odd-parity charge transfer exciton. This local coupling stabilizes the $2A_g$ state and induces a nearest neighbor attraction between the triplets. In addition, because of the electron-hole attraction in the exciton, the increased probability that the electron-hole pair occupies neighboring dimers enhances the triplet-triplet attraction: the triplet pair is `slaved' to  the charge-transfer exciton. The theory also predicts that as the Coulomb interaction increases, the charge-transfer exciton becomes energetically less stable relative to the triplet pair causing a decoupling of the triplet-pair and exciton subspaces. This decoupling has been observed in computational studies of the PPPP model\cite{Book}. Above this Coulomb interaction the triplet-pair binding energy vanishes and the  $2A_g$ state is entirely composed of triplet-pairs.

Triplet-pair interactions within  a dimerized antiferromagnetic chain have been investigated in refs\cite{Harris1973,Uhrig1996,Zheng2001}, while
stabilization of the $2A_g$ state via configuration interactions was also discussed in ref\cite{Taffet20}.

The next section describes the model, while the results are presented in Section III.

\vfill\pagebreak

\section{Model}\label{Se:2}

\subsection{Pariser-Parr-Pople model of $\pi$-conjugated polymers}

Our starting point for a derivation of a low-energy effective model for the $2A_g$ state is the Pariser-Parr-Pople (PPP) model of $\pi$-conjugated polymers.
This model is defined as
\begin{widetext}
\begin{eqnarray}
\hat{H}_{\textrm{PPP}} &=& -\sum_{m\sigma} t_m \left( \hat{c}^{\dag}_{m \sigma}  \hat{c}_{m + 1  \sigma} +  \hat{c}^{\dag}_{m+ 1  \sigma}  \hat{c}_{m   \sigma} \right)
\nonumber\\
& +&
 U \sum_{m}  \big( \hat{N}_{m \uparrow} - \frac{1}{2} \big) \big( \hat{N}_{m \downarrow} - \frac{1}{2} \big)
+  \sum_{m} \sum_{n \ge 1} V_n \big( \hat{N}_m -1 \big) \big( \hat{N}_{m+n} -1 \big),
\end{eqnarray}
\end{widetext}
where $ \hat{c}^{\dag}_{m \sigma}$  ($\hat{c}_{m  \sigma}$) creates (destroys) an electron with spin $\sigma$ in the $p_z$ orbital of carbon atom $m$. $\hat{N}_{m \sigma}$ is the corresponding number operator and $\hat{N}_{m}=\sum_{\sigma}\hat{N}_{m \sigma}$.
Assuming periodic bond alternation, the nearest neighbor electron transfer integral is
\begin{equation}\label{}
  t_m = t_0(1+(-1)^m\delta),
\end{equation}
where $\delta$ is the bond alternation parameter. For convenience, we define $t_d = t_0(1+\delta)$ and  $t_s = t_0(1-\delta)$ as the double and single bond transfer integrals, respectively.
The Coulomb interaction is represented by the Ohno potential, i.e.,
\begin{equation}\label{}
  V_n = \frac{U}{\left( 1+ (U\epsilon r_n/14.397)\right)^{1/2}},
\end{equation}
where $U$ is in eV, the separation between atoms $m$ and $m+n$, $r_n$, is in \AA\ and $\epsilon$ is the relative permittivity.

Typical parameter values for $\pi$-conjugated polymers\cite{Chandross97} are $U=8$ eV, $\epsilon = 2$, $t_0 = 2.4$ eV. In addition, for polyenes, $\delta = 1/12$, so that $t_d = 2.6$ eV and $t_s = 2.2$ eV. In this paper, however, $U$ and $\delta$ are generally arbitrary parameters.

\subsection{Effective model of the $2A_g$ state}\label{Se:2.2}

\begin{figure}
\includegraphics[width=0.6\linewidth]{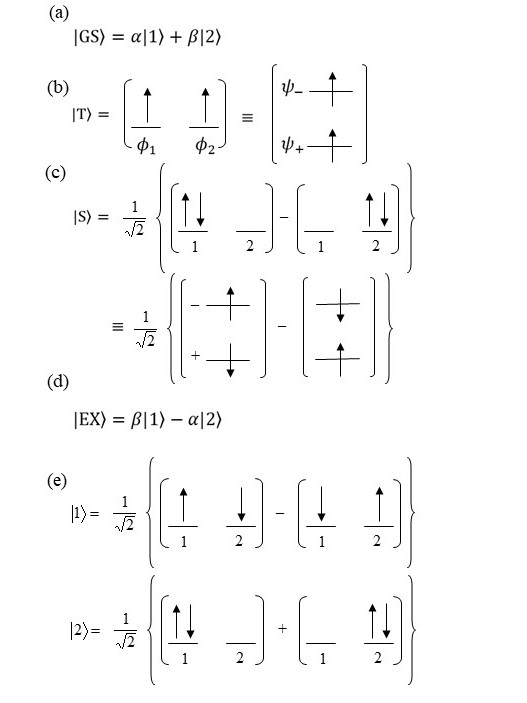}
\caption{The four electronic eigenstates of the two-electron ethylene dimer, (a) $|\textrm{GS}\rangle$, (b) $|\textrm{T}\rangle$, (c) $|\textrm{S}\rangle$  and (d) $|\textrm{EX}\rangle$.
(e) The basis states $|1\rangle$ and $|2\rangle$. The eigenstate energies,  and $\alpha$ and $\beta$ are given in Appendix A.
 Also shown in (b) is the equivalence between the dimer atomic-orbital representation (i.e., $\phi_1$ and $\phi_2$ )  and molecular-orbital representation (i.e., $\psi_{\pm} = (\phi_1 \pm \phi_2)/\sqrt{2}$).}
\label{Fi:1}
\end{figure}

The model assumes that carotenoids and polyenes are linear chains of weakly coupled ethylene dimers, where a dimer is composed of two carbon atoms each with a single $p_z$ orbital. The four electronic states of a dimer with two $\pi$-electrons are illustrated in Fig.\ \ref{Fi:1}. The singlet ground state, $|\textrm{GS}\rangle = \alpha|1\rangle + \beta|2\rangle$, is a linear combination of covalent  and ionic  states (i.e.,  $|1\rangle$ and  $|2\rangle$). As shown by Eq.\ (\ref{Eq:A6}), for noninteracting electrons, $\alpha = \beta = 1/\sqrt{2}$, but as $U/t_d \rightarrow \infty$, $\alpha \rightarrow 1$. For realistic polyene parameters $\alpha \sim \sqrt{2/3}$.

\begin{figure}
\includegraphics[width=0.8\linewidth]{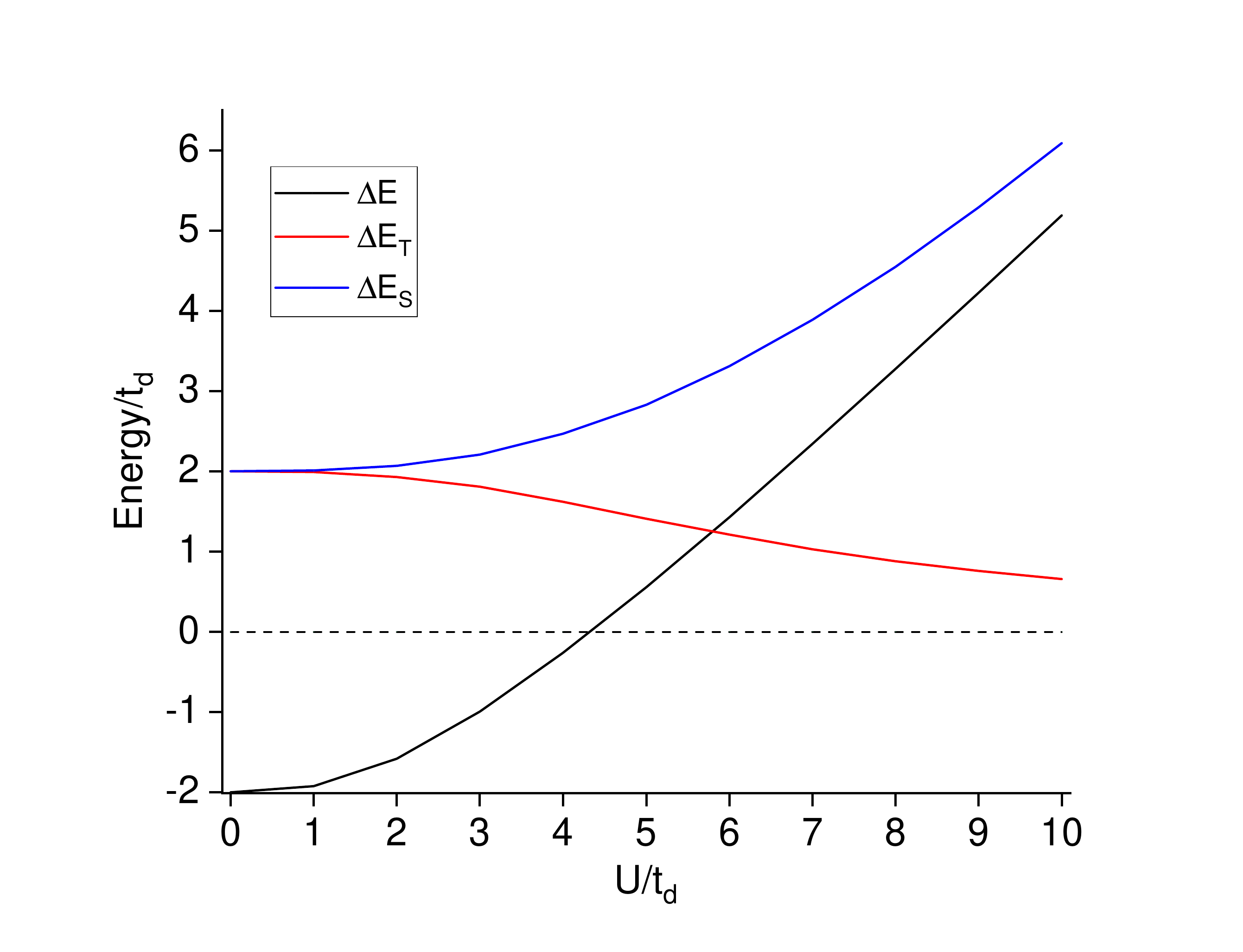}
\caption{The ethylene dimer triplet, $\Delta E_\textrm{T}$, and singlet, $\Delta E_\textrm{S}$, excitation energies as a function of $U/t_d$.
Also shown for neighboring  dimers is the  energy of the charge-transfer exciton relative to a pair of triplets, $\Delta E = ((E_\textrm{CT}-\tilde{V}_1) - E_\textrm{TT})$, where $-\tilde{V}_1$ is the nearest neighbor electron-hole interaction.}
\label{Fi:2}
\end{figure}

In general, the lowest energy excitation is the covalent triplet state, $|\textrm{T}\rangle$, while the lowest singlet excitation is the ionic state, $|\textrm{S}\rangle$. The excitation energies of these two states as a function of $U/t_d$ are illustrated in Fig.\ \ref{Fi:2}. In the noninteracting limit (i.e., $U=0$) these excitation energies are degenerate at $2t_d$. However, as $U$ increases, the ground state becomes more covalent and the triplet excitation energy decreases as $4t_d^2/(U-V_1)$. In contrast, the singlet excitation energy increases as $(U-V_1)$.
Appendix A discusses the dimer solutions in more detail.

\subsubsection{Singlet triplet-pair basis}

We now make the assumption  that the ground state of polyenes can be approximated as a product of dimer ground states, i.e.,
 $ |\textrm{GS}\rangle = \Pi_i |\textrm{GS}\rangle_i$,
where the product is over all dimers and $|\textrm{GS}\rangle_i$ is the  ground state for dimer $i$, defined in Eq.\ (\ref{Eq:A3}).

A triplet excitation from the polyene ground state then corresponds to a triplet excitation on a single dimer (i.e., $|\textrm{T}\rangle$). We denote this state as  $|S=1,M_S;i\rangle$, where $M_S$ is the spin-projection (i.e., $1$, $0$, or $-1$) and $i$ labels the dimer. A triplet on dimer $i$ can hop to a neighboring dimer in its singlet ground state by a superexchange mechanism, i.e., via a virtual charge-transfer state  higher in energy by $(U-V_1)$. By second order perturbation theory, the superexchange transfer integral is
\begin{equation}\label{Eq:2}
  t_{\textrm{TT}} = -\alpha^2 \frac{t_s^2}{(U-V_1)},
\end{equation}
where $\alpha^2$ (defined in Eq.\ (\ref{Eq:A6})) is the probability that the neighboring singlet dimer is in the covalent state labeled $|1\rangle$ in Fig.\ \ref{Fi:1}.

\begin{figure}
\includegraphics[width=0.8\linewidth]{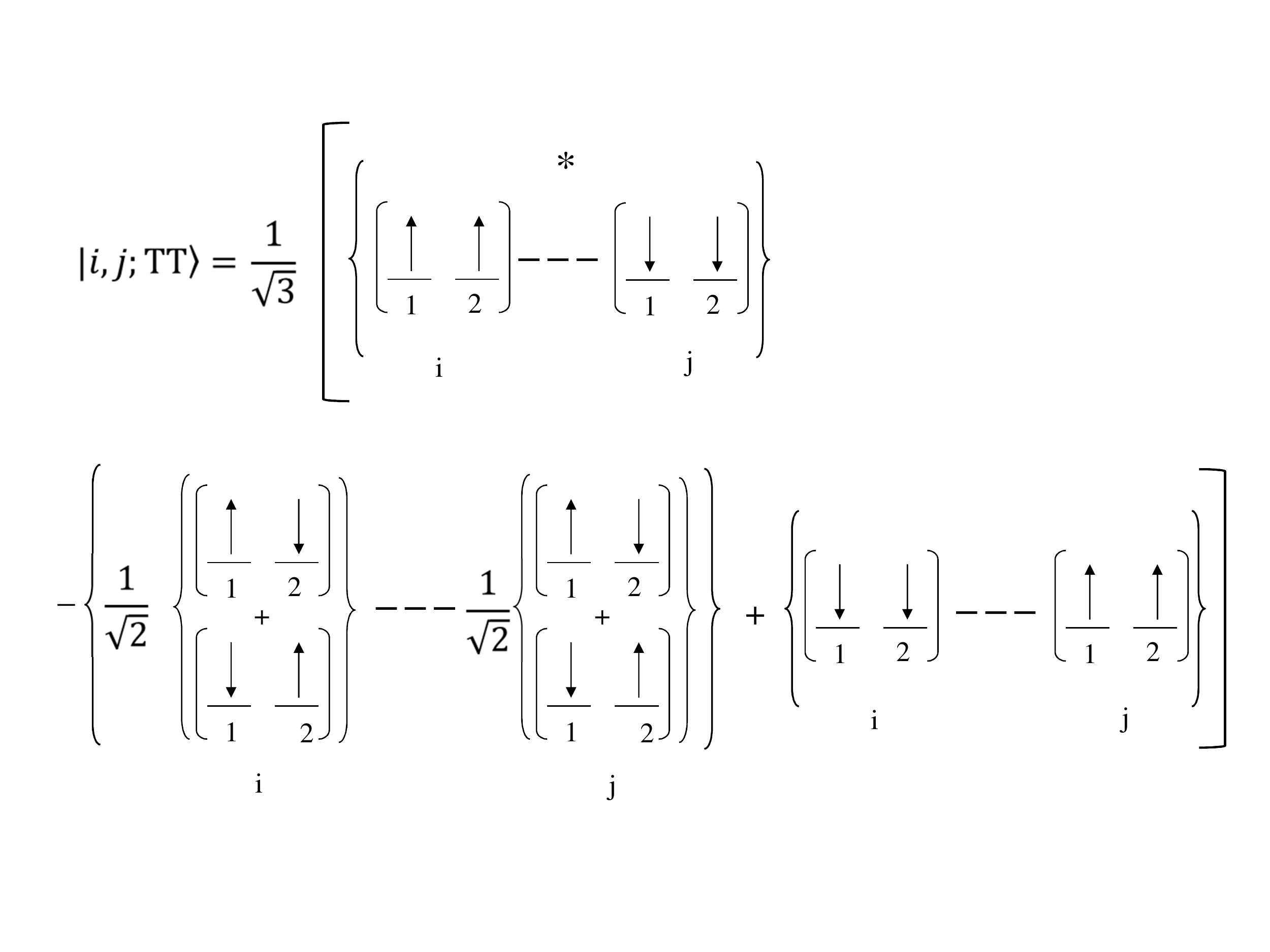}
\caption{Atomic orbital representation of the singlet triplet-pair basis state, \\
$ |i,j;\textrm{TT}\rangle = \left( |1,1;i\rangle|1,-1;j\rangle - |1,0;i\rangle|1,0;j\rangle + |1,-1;i\rangle|1,1;j\rangle \right)/\sqrt{3}$,
with triplet excitations on dimers $i$ and $j$. The dashed lines $ - - -$ represent $(j-i-1)$ dimers in their singlet ground state.
When $j=i+1$ the state labeled $*$ is connected to the state labeled $\times$ in Fig.\ \ref{Fi:4} via an electron transfer across the single bond.}
\label{Fi:3}
\end{figure}

A triplet-pair excitation corresponds to  excitations on separate dimers, $i$ and $j$. A pair of spin-correlated triplets can form an overall singlet, triplet or quintet state. Here, we concerned with the singlet triplet-pair state, expressed as
\begin{widetext}
\begin{equation}\label{}
  |i,j;\textrm{TT}\rangle = \frac{1}{\sqrt{3}}\left( |1,1;i\rangle|1,-1;j\rangle - |1,0;i\rangle|1,0;j\rangle + |1,-1;i\rangle|1,1;j\rangle \right)
\end{equation}
\end{widetext}
and illustrated schematically in Fig.\ \ref{Fi:3}. 

Thus, the Hamiltonian that describes the singlet triplet-pair subspace is
\begin{widetext}
\begin{eqnarray}
  \hat{H}_{\textrm{TT}} &&=
  E_{\textrm{TT}}\sum_{i=1}^{N-1}\sum_{j=i+1}^{N}  \left|i,j;\textrm{TT}\rangle \langle i,j;\textrm{TT}\right|
\nonumber \\
&& +  t_{\textrm{TT}}\left[ \sum_{i=2}^{N-1}\sum_{j=i+1}^{N}\left(\left|i-1,j;\textrm{TT}\rangle \langle i,j;\textrm{TT}\right| + \textrm{H.C.} \right)
+
\sum_{i=1}^{N-2}\sum_{j=i+1}^{N-1}\left(\left|i,j+1;\textrm{TT}\rangle \langle i,j;\textrm{TT}\right| + \textrm{H.C.} \right)
\right].
\nonumber \\
\end{eqnarray}
\end{widetext}
The first term on the right-hand-side describes the energy to excite a pair of triplets on different dimers $i$ and $j$, where
\begin{equation}\label{Eq:5a}
  E_{\textrm{TT}} = 2V_1-2 E_{\textrm{GS}}^{\textrm{dimer}}
\end{equation}
and $E_{\textrm{GS}}^{\textrm{dimer}}$ is defined in Eq.\ (\ref{Eq:A4}).
The second pair of terms describes the hopping of each triplet onto neighboring singlet dimers, while avoiding a hop onto the same dimer.

\subsubsection{Charge-transfer basis}

\begin{figure}
\includegraphics[width=0.8\linewidth]{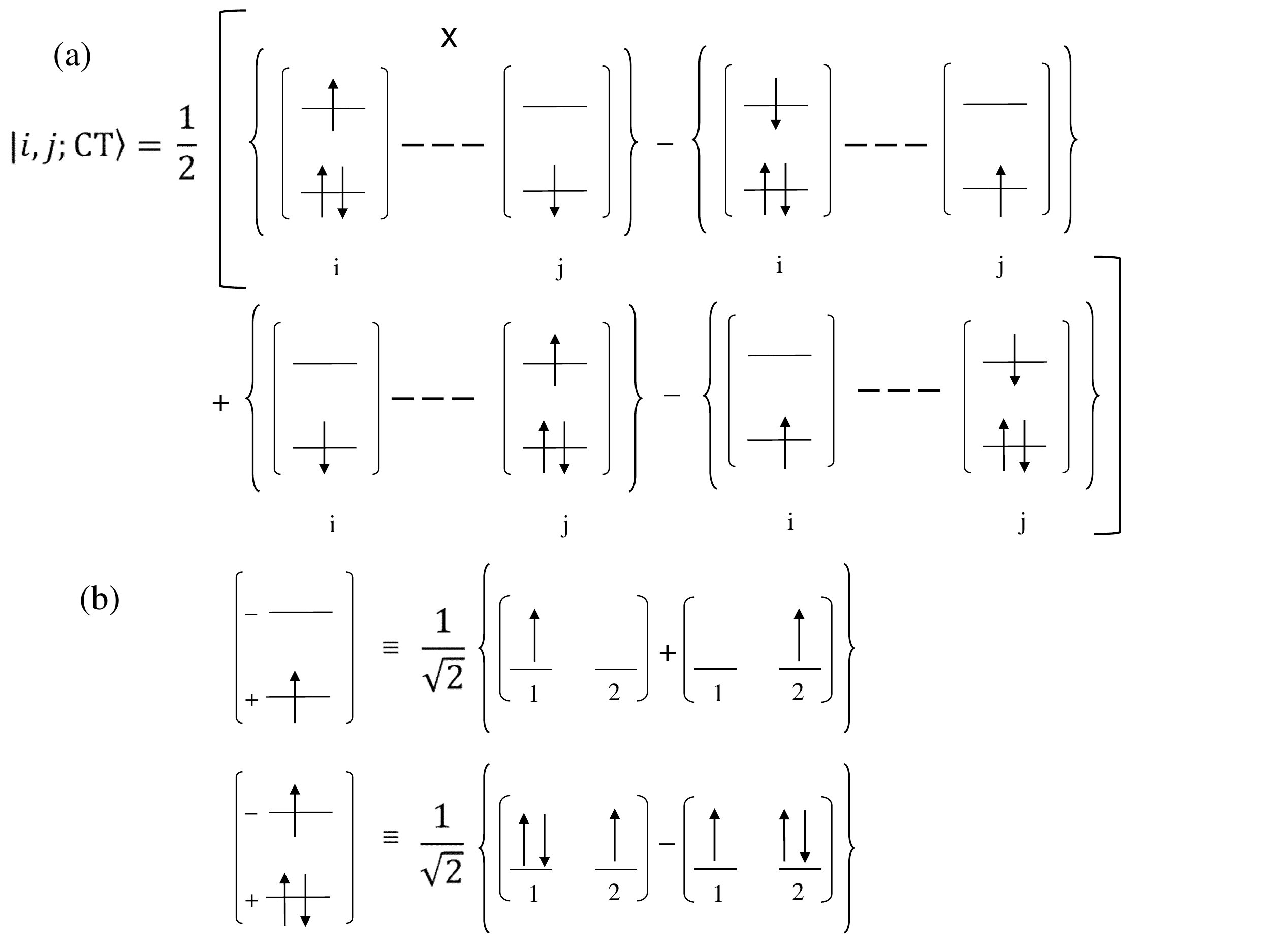}
\caption{(a) Molecular orbital representation of the odd-parity singlet charge-transfer basis state, $ |i,j;\textrm{CT}\rangle$, with electron-hole excitations on dimers $i$ and $j$.
(b) The equivalence between the dimer molecular-orbital and atomic-orbital representations.
When $j=i+1$ the state labeled $\times$ is connected to the state labeled $*$ in Fig.\ \ref{Fi:3} via an electron transfer across the single bond.}
\label{Fi:4}
\end{figure}

The second kind of excitation from the ground state are electron-hole excitations. The dimer state  $|\textrm{S}\rangle$ shown in Fig.\ \ref{Fi:1} is the basis state for the tightly bound Frenkel exciton of conjugated polymers\cite{Book}. Because of its even electron-hole parity, however, this state cannot couple with a singlet triplet-pair state. For an electron-hole excitation to couple to a triplet-pair state it must have odd electron-hole parity, which implies a charge-transfer state as illustrated in Fig.\ \ref{Fi:4}.
The electron in the antibonding $\psi_-$ dimer orbital and a hole in the bonding $\psi_+$ dimer orbital hop between neighboring dimers via the transfer integral
\begin{equation}\label{Eq:6}
  t_{\textrm{CT}} = t_s/2,
\end{equation}
where the factor of $1/2$ arises from the overlap of neighboring dimer orbitals.

The Hamiltonian that describes the odd-parity electron-hole subspace is
\begin{widetext}
\begin{eqnarray}\label{Eq:5}
  \hat{H}_{\textrm{CT}} &&=
 \sum_{i=1}^{N-1}\sum_{j=i+1}^{N} (E_{\textrm{CT}} - \tilde{V}_{(j-i)}) \left|i,j;\textrm{CT}\rangle \langle i,j;\textrm{CT}\right|
\nonumber
\\
&& +  t_{\textrm{CT}}\left[ \sum_{i=2}^{N-1}\sum_{j=i+1}^{N}\left(\left|i-1,j;\textrm{CT}\rangle \langle i,j;\textrm{CT}\right| + \textrm{H.C.} \right)
+
\sum_{i=1}^{N-2}\sum_{j=i+1}^{N-1}\left(\left|i,j+1;\textrm{CT}\rangle \langle i,j;\textrm{CT}\right| + \textrm{H.C.} \right)
\right]
\nonumber
 \\
\end{eqnarray}
\end{widetext}
The first term on the right-hand-side describes the energy to excite an electron-hole pair  on different dimers $i$ and $j$, where
\begin{equation}\label{Eq:8}
  E_{\textrm{CT}} = (U+2V_1-2t_d))-2 E_{\textrm{GS}}^{\textrm{dimer}}.
\end{equation}
In addition, there is a Coulomb attraction between the electron-hole pair $\ell = (j-i)$ dimers apart, given by
\begin{equation}\label{Eq:9}
  \tilde{V}_{\ell} = (V_{2\ell -1}+2V_{2\ell}+V_{2\ell +1})/4.
\end{equation}
The final two terms describe the (symmetrized) motion of the electron or hole between neighboring dimers.

As described in ref\cite{Barford02a}, in the continuum limit for a $1/r$ Coulomb interaction
there is a Rydberg series of  odd-parity bound electron-hole  pairs. The $2A_g$ charge-transfer exciton state is the lowest energy member of this series.\footnote{The even-parity excitons alternate in energy with the odd-parity excitons, while the lowest even-parity (Frenkel) exciton is split off from the Rydberg series\cite{Barford02a,Book}.}

The relative energies of the triplet-pair  and electron-hole pair is illustrated in Fig.\ \ref{Fi:2}. In the noninteracting limit a single triplet is degenerate with an electron-hole excitation, so $\Delta E = (E_{\textrm{CT}}-E_{\textrm{TT}})$ is negative. However, as we have seen, as $U$ increases the triplet energy decreases while the electron-hole energy increases, causing a reversal of energies at $U \sim 4t_d$. It is this property of covalent and ionic states which partially stabilizes the $2^1A_g^-$ state relative to the optically active $1^1B_u^+$ Frenkel exciton state in polyenes.

\subsubsection{Triplet-pair and charge-transfer pair coupling}

We now consider the coupling between the singlet triplet-pair and the odd-parity charge-transfer subspaces. For the special case that a pair of triplet excitations occupy neighboring dimers, i.e., $|i,i + 1;\textrm{TT}\rangle$, an electron transfer across the single bond between them creates a nearest neighbor electron-hole pair, i.e.,  $|i,i+ 1;\textrm{CT}\rangle$. This may be understood by examining the basis states labeled $*$ in Fig.\ \ref{Fi:3} and  $\times$ Fig.\ \ref{Fi:4}. By inspection of $*$ in Fig.\ \ref{Fi:3}, we observe that a transfer of the down electron on site $1$ of dimer $j=i+1$ to site $2$ of dimer $i$ creates a component of the basis state $\times$ in Fig.\ \ref{Fi:4}.

The Hamiltonian describing this process is
\begin{widetext}
\begin{eqnarray}
  \hat{H}_{\textrm{CT-TT}} =
-V_{\textrm{CT-TT}} \sum_{i=1}^{N-1} \left(\left|i,i+1;\textrm{CT}\rangle \langle i,i+1;\textrm{TT}\right| + \left|i,i+1;\textrm{TT}\rangle \langle i,i+1;\textrm{CT}\right| \right),
\end{eqnarray}
\end{widetext}
where
\begin{equation}\label{Eq:11}
V_{\textrm{CT-TT}} = \sqrt{3} t_s/2.
\end{equation}
Since this term only connects a nearest-neighbor triplet pair with a nearest-neighbor electron-hole pair, it results in an attraction between the triplet pair  and an additional attraction above the Coulomb interaction for the electron-hole pair.

\subsubsection{The $2A_g$ state}

We express an eigenstate of the full effective Hamiltonian,
  $\hat{H} = \hat{H}_{\textrm{TT}}+\hat{H}_{\textrm{CT}}+\hat{H}_{\textrm{CT-TT}},$
as
\begin{eqnarray}\label{Eq:}
|\Psi\rangle = \sum_{i, j>i} \Psi_{ij}^{\textrm{TT}} |i,j;\textrm{TT}\rangle + \Psi_{ij}^{\textrm{CT}} |i,j;\textrm{CT}\rangle.
\end{eqnarray}
Since this is a linear combination of basis states from the triplet-pair and electron-hole pair subbases, and not formed via a direct product of the subbases, the interaction $V_{\textrm{CT-TT}}$ does not correlate (or entangle) the triplet-pair and charge-transfer exciton. Instead, it stablizes the linear combination and causes a triplet-triplet attraction.

\begin{figure}[h]
\includegraphics[width=0.6\linewidth]{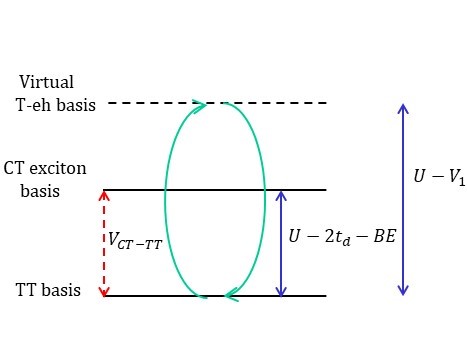}
\caption{A schematic diagram of the relative energy levels of the TT subspace, the CT exciton subspace, and the virtual triplet-electron-hole subspace by which a triplet hops to a neighboring dimer in its singlet ground state (indicated by green arrows). The coupling of the TT  and CT exciton subspaces is indicated by the red-dashed arrow, which only occurs when the triplet pair occupies neighboring dimers.  $BE$ is the CT exciton binding energy.}
\label{Fi:5}
\end{figure}
Before discussing the predictions of this model, we conclude this section by illustrating the separation of energy scales via Fig.\ \ref{Fi:5}. This shows (for relevant polyene parameters) the triplet-pair subspace lying lower in energy than the charge-transfer exciton subspace. Lying higher still in energy is the virtual triplet-electron-hole subspace, through which the triplets delocalize via a second-order process.

\vfill\pagebreak

\section{Results and Discussion}\label{Se:3}

We now describe the solutions of the effective model of the $2A_g$ state. We begin by exploring the parameter regime with arbitrary parameters for a translationally invariant system of 1000 monomers. We define two energy gaps. First, the stabilization energy gained by the $2A_g$ state caused by the coupling of the triplet and electron-hole pair subspaces, i.e.,
\begin{equation}\label{Eq:12}
 \Delta_{\textrm{CT-TT}} =   E(V_{\textrm{CT-TT}}=0)-E(V_{\textrm{CT-TT}}).
\end{equation}
Second, the triplet-triplet pair binding energy, defined as the energy gap between the band of free triplet-pairs and the $2A_g$ state, i.e.,
\begin{equation}\label{Eq:13}
 \Delta_{\textrm{TT}} =  (E_{\textrm{TT}}-4|t_{\textrm{TT}}|)-E(V_{\textrm{CT-TT}}).
\end{equation}
The state will be characterized by its triplet-pair weight, i.e.,
\begin{equation}\label{Eq:14}
  P_{\textrm{TT}} = \sum_{i, j>i} \left|\Psi_{ij}^{\textrm{TT}}\right|^2,
\end{equation}
and the mean triplet-pair separation (in monomer units), i.e.,
\begin{equation}\label{Eq:15}
  L_{\textrm{TT}} = \frac{\sum_{i, j>i} (j-i)\left|\Psi_{ij}^{\textrm{TT}}\right|^2}{P_{\textrm{TT}}}.
\end{equation}

\subsection{Toy Model Parameters}

We begin our investigation of the predictions of the model by considering the simplest limit of non-interacting electron-hole pairs, i.e., we set the Coulomb interaction $\tilde{V}_{(j-i)} = 0$ in Eq.\ (\ref{Eq:5}). In this limit there is a symmetry between the triplet-pair and the electron-hole pair under the exchange $E_{\textrm{CT}} \leftrightarrow E_{\textrm{TT}}$ and $|t_{\textrm{CT}}| \leftrightarrow |t_{\textrm{TT}}|$. In the absence of the coupling, $V_{\textrm{CT-TT}}$, between the two subspaces the triplet-pair forms a band of free triplet excitations (subject to a hard-core repulsion) centered at $E_{\textrm{TT}}$ of width $8|t_{\textrm{TT}}|$. Likewise, there is a band of unbound electron-hole excitations centered at $E_{\textrm{CT}}$ of width $8|t_{\textrm{CT}}|$.

Turning on the coupling, $V_{\textrm{CT-TT}}$, causes the nearest neighbor triplet-pair to mix with the nearest neighbor electron-hole pair. At resonance, i.e., when $E_{\textrm{CT}} = E_{\textrm{TT}}$ and $|t_{\textrm{CT}}| = |t_{\textrm{TT}}|$, the mixing causes a nearest neighbor attraction of $V_{\textrm{CT-TT}}$ for both the triplet and electron-hole pairs. This model then maps onto the well-known one-dimensional lattice model of spinless fermions interacting with a nearest neighbor attraction, $V_{\textrm{CT-TT}}$, for which above a critical value of $V_{\textrm{CT-TT}}$ there is a single band of bound states. As shown in refs\cite{Mattis1988,Gallagher1997,Gebhard1997}, the binding energy
for the zero-momentum state  is
\begin{equation}\label{Eq:7}
E_{BE} =  V_{\textrm{CT-TT}} +\frac{4 t_{\textrm{TT}}^2}{V_{\textrm{CT-TT}}}
 -     4|t_{\textrm{TT}}|,
\end{equation}
implying a critical value of $V_{\textrm{CT-TT}}$ of $2|t_{\textrm{TT}}|$.

\begin{figure}
\includegraphics[width=0.8\linewidth]{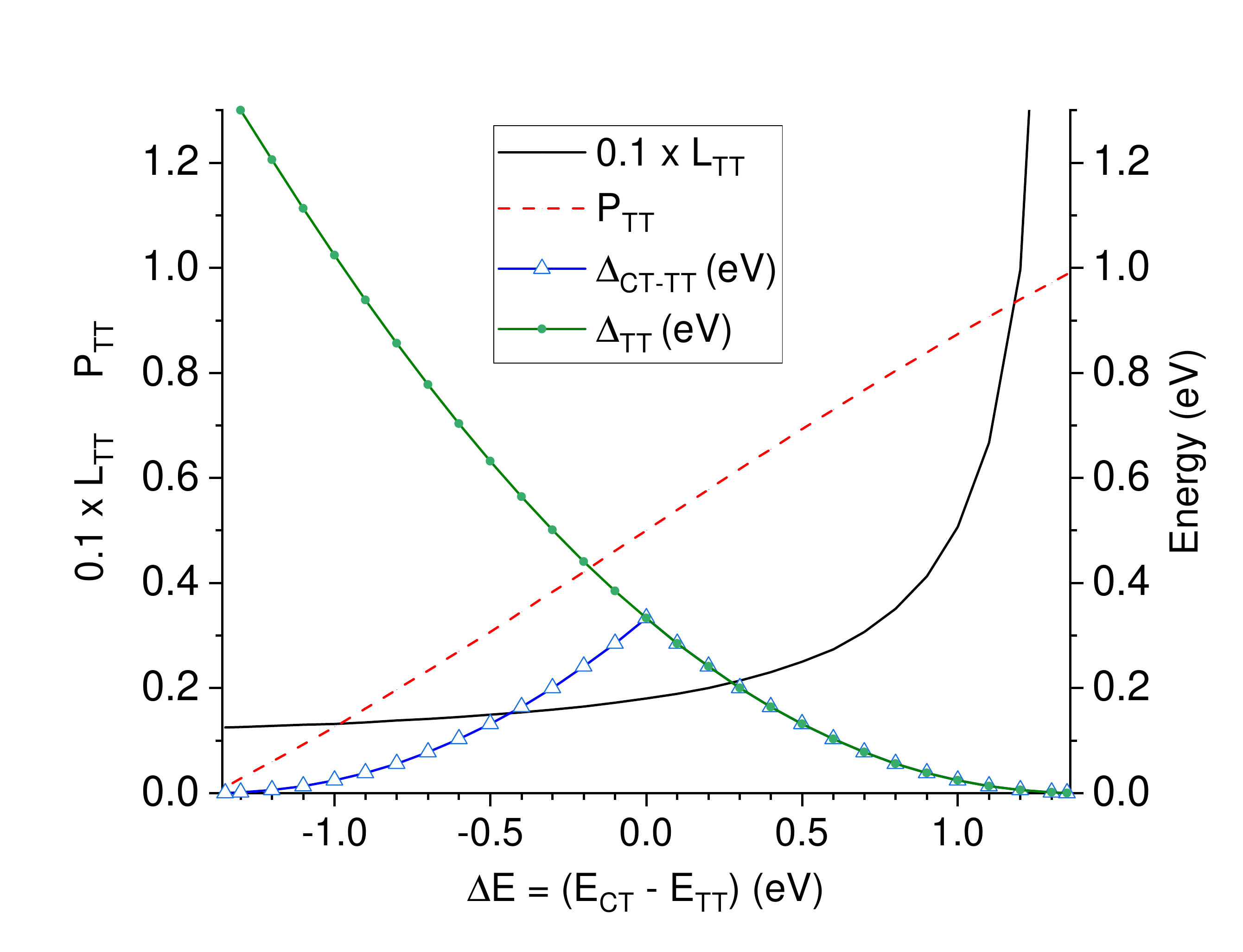}
\caption{Non-Coulombically interacting electron-hole pairs (i.e., $U=0$).
$\Delta_{\textrm{CT-TT}}$, Eq.\ (\ref{Eq:12}), is the stabilization energy of the $2A_g$ state caused by the coupling of the triplet-pair and electron-hole pair; $\Delta_{\textrm{TT}}$, Eq.\ (\ref{Eq:13}), is the triplet-pair binding energy; $L_{\textrm{TT}}$, Eq.\ (\ref{Eq:14}), is the mean triplet-pair separation (scaled by $0.1$ and in units of dimers); and $P_{\textrm{TT}}$, Eq.\ (\ref{Eq:15}), is the triplet-pair weight.
$\Delta_{\textrm{TT}}=\Delta_{\textrm{CT-TT}}$ when $\Delta E \ge 0$.
$t_{TT} = -t_{CT} = -1.0$ eV, $V_{CT-TT}  = 3.0$ eV.
$\Delta E_{\textrm{critical}} = \pm 1.35$ eV, above and below which the triplet-pair and electron-hole pair subspaces are decoupled.
}
\label{Fi:6}
\end{figure}

Figure \ref{Fi:6} illustrates the solutions when $|t_{\textrm{CT}}| = |t_{\textrm{TT}}|=1$  eV and $V_{\textrm{CT-TT}}=3$ eV. At resonance (i.e., $\Delta E = (E_{\textrm{CT}}-E_{\textrm{TT}})= 0$), the $2A_g$ state, $|\Psi\rangle$, consists of an equal mixture of bound triplet pairs and bound electron-holes pairs. The mean triplet-triplet and electron-hole separation in both pairs is ca.\ 2 dimers. The energy gap between this state and the bands of free triplets and electron-holes (i.e., $\Delta_{\textrm{CT-TT}}$) is $|t_{\textrm{TT}}|/3$, in agreement with Eq.\ (\ref{Eq:7}). At resonance this energy gap corresponds to the binding energies of both the triplet-pairs and electron-hole pairs, i.e., $\Delta_{\textrm{CT-TT}} = \Delta_{\textrm{TT}}  =\Delta_{\textrm{CT}}$.

As the energy of the  band of electron-hole excitations increases, i.e., as $E_{\textrm{CT}}$ increases, the system goes off resonance and the $2A_g$ state  is predominately composed of bound triplet-pairs; the triplet-pair binding energy decreases and the  mean triplet-pair separation increases. At a critical value of $\Delta E$, namely $\Delta E_{\textrm{critical}}^{\textrm{U}} = 1.35$ eV, there is a decoupling of the triplet-pair and electron-hole pair subspaces, and the triplet-pair and the electron-hole pair both become unbound. For $\Delta E \ge \Delta E_{\textrm{critical}}^{\textrm{U}} $, $P_{\textrm{TT}} = 1$.

Similarly, as $E_{\textrm{CT}}$ decreases from resonance the $2A_g$ state  is predominately composed of bound electron-hole pairs. Now the triplet-pair binding energy, $\Delta_{\textrm{TT}}$, increases and the  mean triplet-pair separation decreases. Again,  there is  critical value of $\Delta E$, namely $\Delta E_{\textrm{critical}}^{\textrm{L}} = -1.35$ eV, below which  there is a decoupling of the triplet-pair and electron-hole pair subspaces, and both the triplet-pair and electron-hole pair become unbound. For $\Delta E \le \Delta E_{\textrm{critical}}^{\textrm{L}} $, $P_{\textrm{TT}} = 0$.

\begin{figure}
\includegraphics[width=0.8\linewidth]{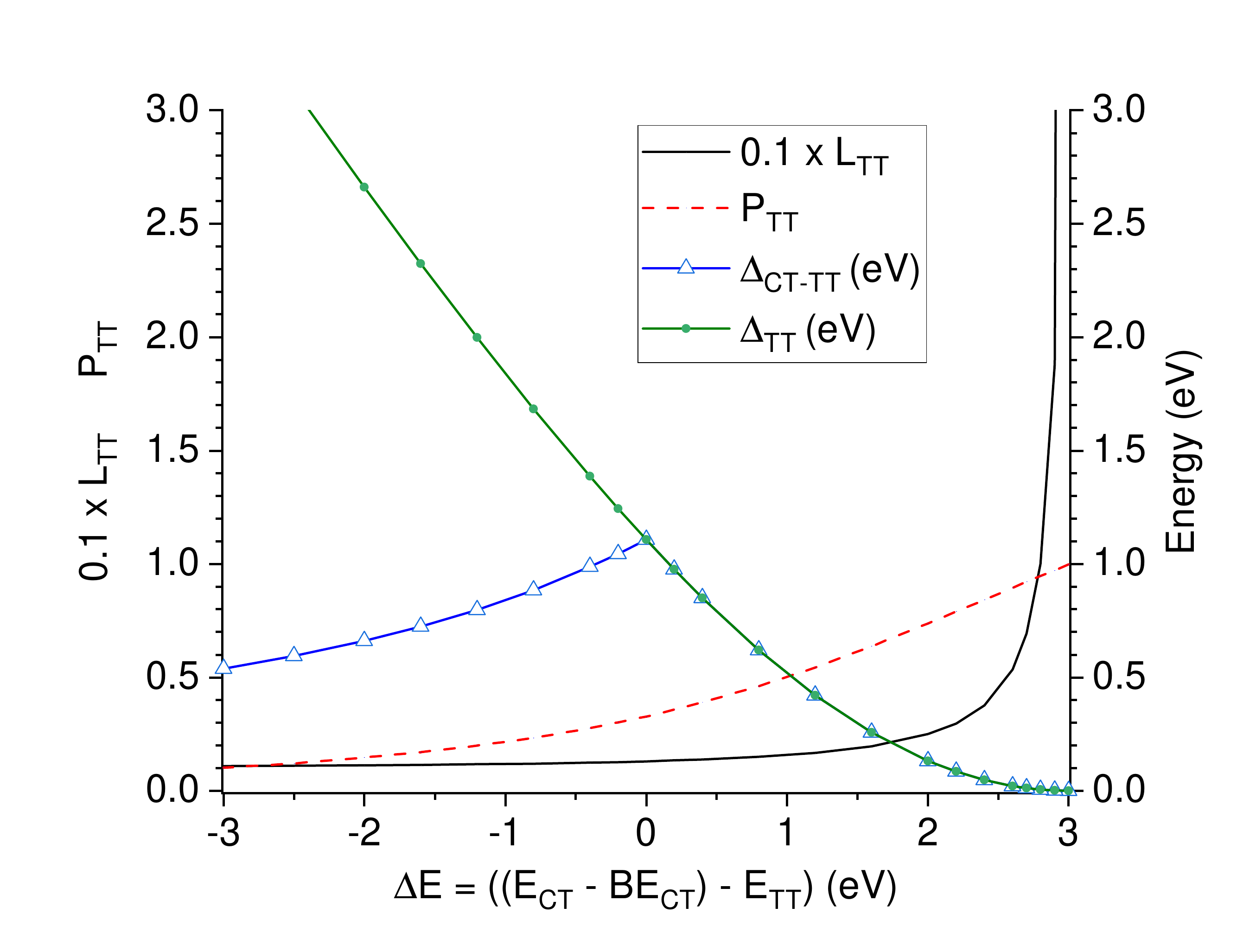}
\caption{Coulombically interacting electron-hole pairs ($U=6$ eV). $t_{TT} = -t_{CT} = -1.0$ eV, $V_{CT-TT}  = 3.0$ eV. $\Delta E_{\textrm{critical}} = +3.0$ eV above  which the triplet-pair and electron-hole pair subspaces are decoupled.
The charge-transfer exciton binding energy, $BE_{\textrm{CT}} = 0.74$ eV.}
\label{Fi:7}
\end{figure}

We now turn to the physical limit of interacting electron-hole pairs, i.e., we set the Coulomb interaction $\tilde{V}_{(j-i)} > 0$ in Eq.\ (\ref{Eq:5}). We set $U = 6$ eV, giving a charge-transfer exciton binding energy of $0.74$ eV.
As illustrated in Fig.\ \ref{Fi:7}, the electron-hole pair attraction causes a number of significant changes to the $\tilde{V}_{(j-i)} = 0$ picture. As for Fig.\ \ref{Fi:6}, the results shown are for $|t_{\textrm{CT}}| = |t_{\textrm{TT}}|=1$ eV and $V_{\textrm{CT-TT}}=3$ eV.
At resonance
the triplet-pair and charge-transfer exciton are mutually stabilized by $1.01$ eV. As $1.01$ eV is  the excitation energy  to the free triplet-pair band, this is also the triplet-pair binding energy. This value is considerably enhanced over the $\tilde{V}_{(j-i)} = 0$ limit of $0.333$ eV, as the triplets are `slaved' to the Coulombically bound electron-hole pair. We also note that the triplet contribution is only 33\%, compared to 50\% at resonance in the absence of Coulombically bound electron-hole pairs. Again, as $E_{\textrm{CT}}$ is increased the triplet-pair binding energy decreases, while  the triplet-pair separation and fraction increases. Above a critical value of $E_{\textrm{CT}}$, given by $\Delta E \gtrsim 3.0$ eV, there is a decoupling of the triplet-pair and electron-hole subspaces and the triplet-pair unbinds. In contrast, as $E_{\textrm{CT}}$ is decreased there is no such decoupling of the triplet-pair and electron-hole subspaces. In this case the triplet-pair becomes more tightly bound, albeit its contribution to the $2A_g$ state becomes less than 10\% for $\Delta E < -3.0$ eV.

\subsection{PPP Model Parameters}

\begin{figure}
\includegraphics[width=0.8\linewidth]{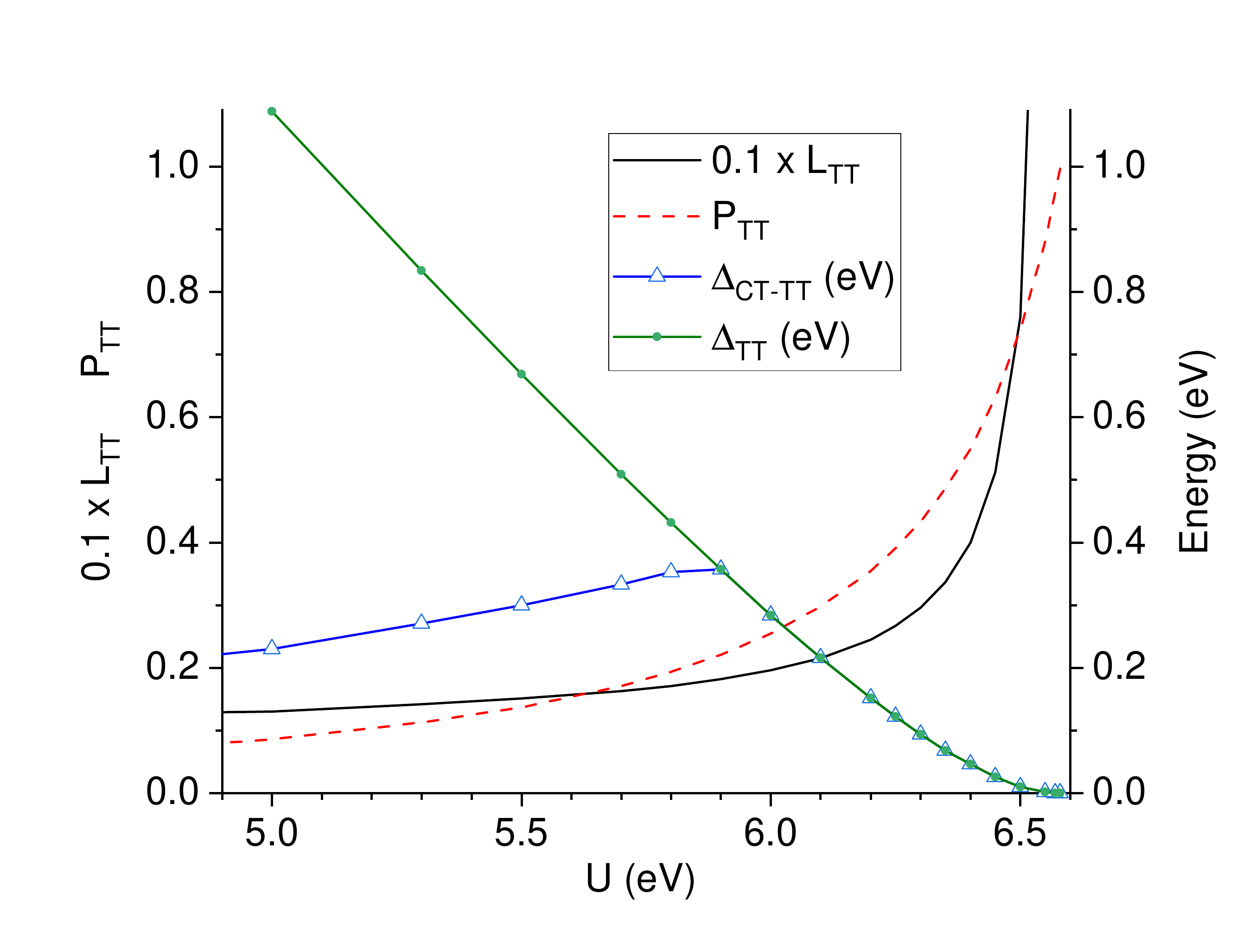}
\caption{Results as a function of the  Coulomb repulsion, $U$, using effective-model parameters derived from the PPP model (see Section \ref{Se:2.2}).
The bond alternation, $\delta = 1/12$.
For $U > 5.9$ eV the triplet-pair state is more stable than the charge-transfer exciton.
The triplet-pair and charge-transfer subspaces decouple for $U \ge 6.6$ eV.}
\label{Fi:8}
\end{figure}

The previous section described the predictions for arbitrary parameters. We now consider the prediction for model parameters derived from the underlying PPP model, Eq.\ (1). In this case,
as shown by Eq.\ (\ref{Eq:2}), Eq.\ (\ref{Eq:5a}), Eq.\ (\ref{Eq:6}), Eq.\ (\ref{Eq:8}), Eq.\ (\ref{Eq:9}) and Eq.\ (\ref{Eq:11}), the model parameters are determined by the Coulomb interaction, $U$, and the bond dimerization, $\delta$.

Figure \ref{Fi:8} illustrates the results by varying the Coulomb interaction, $U$, and for a realistic fixed bond alternation for polyenes, i.e., $\delta = 1/12$.
Since $(E_{\textrm{CT}} - E_{\textrm{TT}}) = (U-2t_d)$, Fig.\ \ref{Fi:8} qualitatively resembles Fig.\ \ref{Fi:7}, namely as $U$ is increased the triplet-pair band becomes more stable relative to the charge-transfer exciton. Resonance (i.e., $\Delta_{\textrm{TT}}= \Delta_{\textrm{CT-TT}}$) occurs at $U = 5.9$ eV, where  $\Delta_{\textrm{TT}} = 0.36$ eV and $P_{\textrm{TT}} = 0.22$.
For  $U \ge 6.6$ eV the triplet-pair and charge-transfer subspaces decouple, implying that the triplet-pair binding energy vanishes and the $2A_g$ state is entirely composed of triplet-pairs. This  prediction of decoupling above a critical $U$ is consistent with DMRG calculations of the PPPP model, as shown in Fig.\ 7.8 of \cite{Book}.
As before, as $U$ is reduced the converse does not happen and the subspaces only decouple at $U=0$.

This  model predicts that for a realistic value of $U = 8$ eV and $\delta = 1/12$, the triplet-pair and charge-transfer subspaces are decoupled. We explain this unphysical prediction as follows. First,  the model is expected to become more exact in the dimer limit, i.e., as $\delta \rightarrow 1$. Indeed, when $\delta = 0.16$, resonance  occurs at $U = 6.4$ eV (where  $\Delta_{\textrm{TT}} = 0.31$ eV and $P_{\textrm{TT}} = 0.24$) and decoupling of the subspaces occurs at $U = 9.9$ eV.
Second, the reduced basis model underscreens Coulomb interactions, so  a smaller value of $U$ is required to reproduce experimental ionic excitation energies.

\begin{figure}
\includegraphics[width=0.8\linewidth]{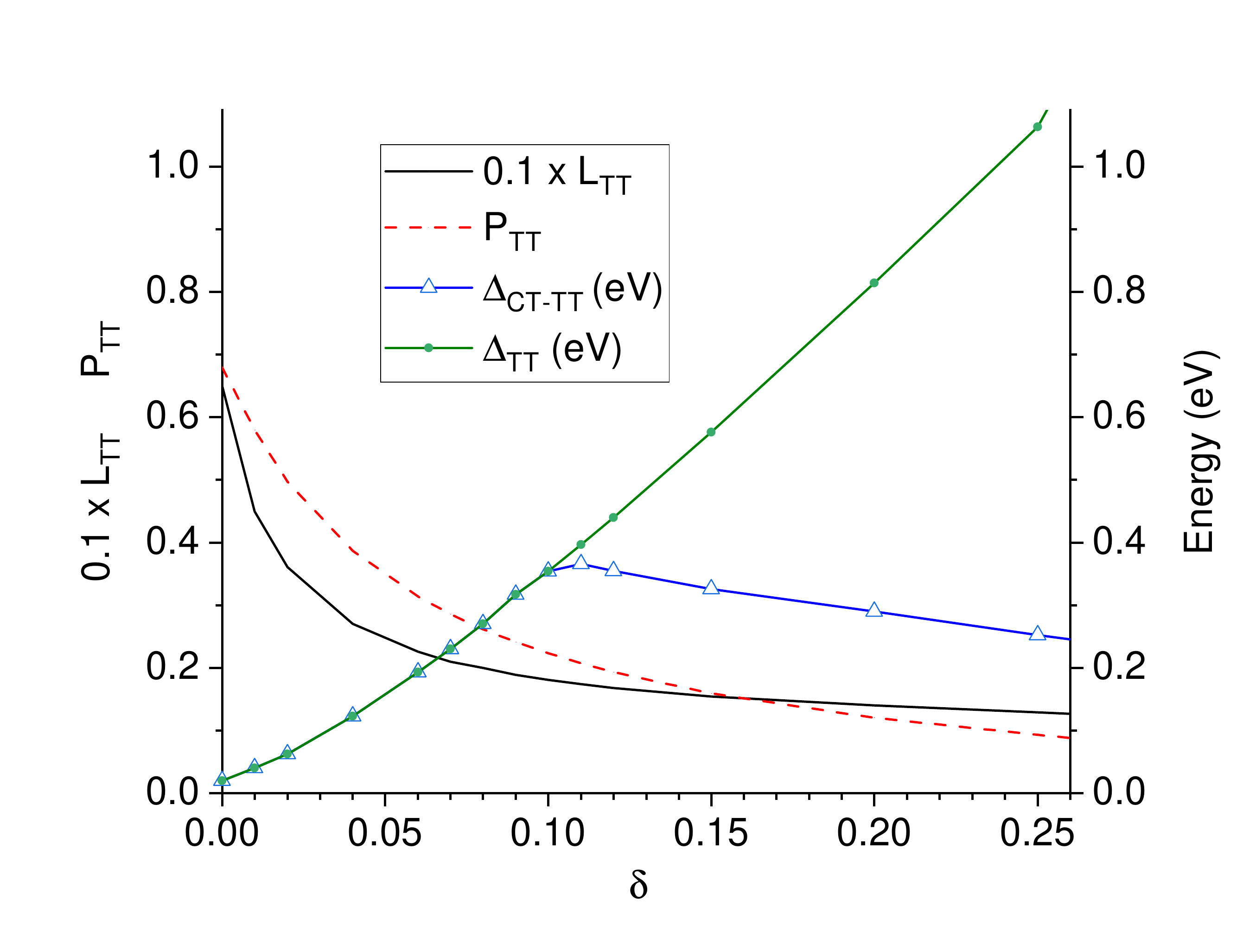}
\caption{Results as a function of the bond alternation, $\delta$, using effective-model parameters derived from the PPP model (see Section \ref{Se:2.2}).
The Coulomb repulsion, $U = 6$ eV.
For  $\delta \le 0.1$  the triplet-pair state is more stable than the charge-transfer exciton.}
\label{Fi:9}
\end{figure}

We explore the role of $\delta$ further via Fig.\ \ref{Fi:9}.
Increasing $\delta$ decreases $E_{\textrm{CT}}$ and hence stabilizes the charge-transfer exciton. It also reduces $t_{\textrm{CT}}$, $|t_{\textrm{TT}}|$ and $V_{\textrm{TT-CT}}$. Thus, the triplet-pair state is more stable than the charge-transfer exciton for  $\delta \le 0.1$. In this case, there is no decoupling of the subspaces.

The results described so far are for the zero-momentum state of a translationally invariant system (with periodic boundary conditions) and 1000 dimers.
We conclude this section by describing our results for carotenoid oligomers (with open boundary conditions) of between 8 and 13 dimers (i.e., 16 and 26 C-atoms). We set $U=6$ eV and $\delta = 1/12$. In this case, compared to the results shown in Fig.\ \ref{Fi:8}, there are some finite-size effects, so the triplet-pair binding energy is increased to $\sim 0.5$ eV. Its weight in the $2A_g$ state is $\sim 25\%$. The binding energy decreases for higher pseudomomentum members of the `$2A_g$' family, but the triplet-pair weight remains the same.

\vfill\pagebreak

\section{Concluding Remarks}\label{Se:4}

This paper has introduced a theory  to describe the singlet dark state (i.e., $S_1$ or  $2A_g$) of polyenes and carotenoids. The theory assumes that in principle this state is a linear combination of a singlet triplet-pair and an odd-parity charge-transfer exciton. Crucially, these components only couple when the triplet-pair occupies neighboring dimers, such that an electron transfer between the triplets creates a nearest-neighbor charge-transfer excitation. This  local coupling stabilises the $2A_g$ state and induces a nearest neighbor attraction between the triplets. In addition, because of the electron-hole attraction in the exciton, the increased probability that the electron-hole pair occupies neighboring dimers enhances the triplet-triplet attraction: the triplet pair is `slaved' to  the charge-transfer exciton.
The reduction of the $2A_g$ energy is an additional cause for the $1^1B_u$/$2^1A_g$ energy reversal in polyenes.

The theory also predicts that as the Coulomb interaction is increased, the $2A_g$ state evolves from a predominately odd-parity charge-transfer exciton state with a small component of triplet-pair character to a state predominately composed of triplet-pairs with some exciton character. Above a critical Coulomb interaction there is a decoupling of the triplet-pair and charge-transfer exciton subspaces, such that the $2A_g$ state becomes entirely composed of an unbound spin-correlated triplet pair.

Although the predictions of this theory are qualitatively consistent with high-level DMRG calculations of the PPP model\cite{Valentine20,Book}, it is not possible to make quantitative predictions for polyenes and carotenoids, because of the approximations of the model. The first approximation is that the  dimer limit is assumed for the ground state and its excitations, implying that the spinons of a triplet are confined to a dimer. Second, the reduced basis of just the triplet-pair and odd-parity electron-hole excitation implies that the Coulomb interactions are underscreened. This means that somewhat smaller values of Coulomb interactions, $U$, or larger values of bond alternation, $\delta$, than are realistic for polyenes are required to obtain semiquantitative agreement with DMRG calculations.



\vfill\pagebreak

\appendix

\section{Dimer Basis States}\label{Se:A1}

This Appendix summarizes the exact results of the PPP model for the eigenstates of the ethylene dimer illustrated in Fig.\ 1.

The triplet excitation energy is
\begin{eqnarray}\label{Eq:A1}
  \Delta E_{\textrm{T}} & & =  V_1 - E_{\textrm{GS}}^{\textrm{dimer}}
  \nonumber\\
  & & \rightarrow 2t_d \textrm{ as }U/t_d \rightarrow 0
    \nonumber\\
  & & \rightarrow  J_d \textrm{ as }U/t_d \rightarrow \infty,
\end{eqnarray}
where $J_d = 4t_d^2/(U-V_1)$,
the ground state energy is
\begin{eqnarray}\label{Eq:A4}
 E_{\textrm{GS}}^{\textrm{dimer}} = (U+V_1)/2 - \epsilon
\end{eqnarray}
and
\begin{equation}\label{}
\epsilon =  \frac{1}{2}\left((U-V_1)^2 + 16 t_d^2\right)^{1/2}.
\end{equation}

The lowest singlet excitation energy is
\begin{eqnarray}\label{Eq:A2}
  \Delta E_{\textrm{S}} & & =  U - E_{\textrm{GS}}^{\textrm{dimer}}
  \nonumber\\
  && \rightarrow 2t_d \textrm{ as }U/t_d \rightarrow 0
    \nonumber\\
  && \rightarrow (U - V_1) + J_d \textrm{ as }U/t_d \rightarrow \infty.
\end{eqnarray}

The ground state is
\begin{equation}\label{Eq:A3}
|\textrm{GS}\rangle = \alpha |1\rangle + \beta |2\rangle,
\end{equation}
where the basis kets $|1\rangle$ and $|2\rangle$ are illustrated in Fig.\ 1, and
\begin{eqnarray}\label{Eq:A6}
  \alpha && =\frac{1}{2} \left(\frac{2\epsilon +(U-V_1)}{\epsilon}\right)^{1/2}
    \nonumber\\
  && \rightarrow \frac{1}{\sqrt{2}} \textrm{ as }U/t_d \rightarrow 0
    \nonumber\\
  && \rightarrow  1 \textrm{ as }U/t_d \rightarrow \infty.
\end{eqnarray}
Also, $\beta = (1-\alpha^2)^{1/2}$.


\vfill\pagebreak
\bibliography{references}

\end{document}